\documentclass[groupedaddress,aps,pra,superscriptaddress,showpacs]{revtex4}%
\usepackage{epsfig,dsfont,amssymb,amsmath,amsthm,amsfonts,amsbsy,mathrsfs}
\usepackage{graphicx, color}
\usepackage{amsmath}
\usepackage{amssymb}
\usepackage{mathdots}
\usepackage{float}
\usepackage{cases}
\usepackage{graphpap}
\usepackage{wrapfig}
\usepackage{tikz}
\usepackage{indentfirst}
\usepackage{picinpar,graphicx}
\DeclareMathOperator{\Tr}{Tr}
\begin{document}
\title{Uncertainty under Quantum Measures and Quantum Memory}
\author{Yunlong Xiao}%mathxiao123@gmail.com
%\thanks{equally contributed to this work }
\affiliation{School of Mathematics, South China University of Technology, Guangzhou, Guangdong 510640, China}
\affiliation{Max Planck Institute for Mathematics in the Sciences, 04103 Leipzig, Germany}
\author{Naihuan Jing\footnote{Corresponding author: jing@ncsu.edu}}
\affiliation{Department of Mathematics, Shanghai University, Shanghai 200444, China}
\affiliation{Department of Mathematics, North Carolina State University, Raleigh, NC 27695, USA}
\author{Xianqing Li-Jost}%xli-jost@mis.mpg.de
\affiliation{Max Planck Institute for Mathematics in the Sciences, 04103 Leipzig, Germany}

%\title{The improved entropic uncertainty principle}

\begin{abstract}
The uncertainty principle restricts potential information one gains about physical properties of the measured particle. However,
if the particle is prepared in entanglement with a quantum memory, the corresponding entropic uncertainty relation will vary. Based on
the knowledge of correlations between the measured particle and quantum memory, we have investigated the entropic uncertainty relations for
two and multiple measurements, and generalized the lower bounds on the sum of Shannon entropies without quantum side
information to those that allow quantum memory. In particular, we have obtained generalization of
Kaniewski-Tomamichel-Wehner's bound for effective measures and majorization bounds for noneffective measures to allow quantum side information. Furthermore, we have derived several strong bounds for the entropic uncertainty relations in the presence of quantum memory for two and
multiple measurements. Finally, potential applications of our results to entanglement witnesses are discussed
via the entropic uncertainty relation in the absence of quantum memory.
\end{abstract}

 %{03.67.-a, 02.20.Hj, 03.65.-w}

%keyowrds: Entropic uncertainty relations, quantum memory, quantum side information

\maketitle

\section{Introduction}

Heisenberg's uncertainty principle \cite{Heisenberg} bounds the limit of measurement outcomes of two incompatible observables,
which reveals a fundamental difference between the classical and quantum mechanics. After intensive studies of the principle
in terms of standard deviations of the measurements, entropies have stood out to be
a natural and important alternative formulation of the uncertainty principle \cite{Marco}. The importance of entropic
uncertainty relations is solidified by a variety of applications, ranging from entanglement witnessing to quantum cryptography.

The first entropic uncertainty relation of observables with finite spectrum
was given by Deutsch \cite{Deutsch} and then improved by Maassen and Uffink \cite{Maassen}, who gave the
celebrated MU bound: if two incompatible measurements $M_{1}=\{|u_{i_{1}}^{1}\rangle\}$ and $M_{2}=\{|u^{2}_{i_{2}}\rangle\}$
are chosen on the particle $A$, then the uncertainty is bounded below by
\begin{align}\label{MU}
H(M_{1})+H(M_{2})\geqslant\log_2\frac{1}{c_{1}},
\end{align}
where $H(M_{i})$ is the {\it Shannon entropy} of the probability distribution
induced by measurement $M_i$ and $c_{1}=\max_{i_{1},i_{2}}\mid\langle u^{1}_{i_{1}}|u^{2}_{i_{2}}\rangle\mid^{2}$
denotes the largest overlap between the observables.
On the other hand,
a mixed state is expected to have more uncertainty, as (\ref{MU}) can be reinforced by adding the complementary term of
the {\it von Neumann entropy}
$H(A)=S(\rho_{A})$:
\begin{align}\label{MUmix}
H(M_{1})+H(M_{2})\geqslant\log_2\frac{1}{c_{1}}+H(A).
\end{align}
The entropy $H(A)$ measures the amount of uncertainty induced by the mixing status of the state $\rho_{A}$:
if the state is pure, then $H(A)=0$, and if the state is a mixed state, then $H(A)>0$.
Therefore the corresponding bound (\ref{MUmix}) is stronger than (\ref{MU})
even though there is no auxiliary quantum system such as a quantum memory. We refer to $\log_2\frac{1}{c_{1}}$ as the classical
part $B_{MU}$ and call $H(A)$ the mixing part
of the bound for the entropic uncertainty relation since it measures the mixing status of the particle.

Most of the bounds for entropic uncertainty relations in the absence of quantum memory contain two parts:
(\expandafter{\romannumeral1}) the classical part $B_{C}$, for instance, Maassen and Uffink's bound \cite{Maassen},
Coles and Piani's bound \cite{Coles}, or our recent bound \cite{XJ};
(\expandafter{\romannumeral2}) the mixing part $H(A)$, which describes the information pertaining to the mixing status of the particle $\rho_{A}$.
We note that both the Kaniewski-Tomamichel-Wehner bound \cite{Kaniewski} based
on effective anti-commutator and the direct-sum majorization bound \cite{Rudnicki} only involve with the classical
part and have no mixing parts. For more details, see Sec. \uppercase\expandafter{\romannumeral2}.
Obviously, not all the bounds $B_{C}$ can be generalized to the case with quantum memory by simply adding
an extra term $H(A|B)$. Therefore it is an interesting problem to extend the entropic uncertainty relations in the absence of quantum memory
to those with quantum memory.

In this paper, we will solve the extension problem by answering three questions:
(\expandafter{\romannumeral1}) Can the uncertainty relation in the absence of quantum memory be generalized to the case
with quantum side information ? % i.e. in the presence of quantum memory?
(\expandafter{\romannumeral2}) Are there other indices besides $H(A|B)$ to quantify the amount of entanglement between the measured
particle and quantum memory?
(\expandafter{\romannumeral3}) Can two pairs of observables sharing the same overlaps between bases have different entropic
uncertainty relations?
Besides answering these questions in detail we will give a couple of strong entropic uncertainty relations in the presence of quantum memory.

\section{Generalized Entropic Uncertainty Relations}

Strengthening the bound for the entropic uncertainty relation is an interesting problem arising from
quantum theory. One of the main issues in this direction is how to extend the entropic uncertainty relation to allow
for quantum side information.
Several approaches have been devoted to seek for stronger bounds for the entropic uncertainty relations (e.g.
majorization-based uncertainty relations, direct-sum majorization relations, uncertainty relations based on effective
anti-commutators and so on). However it is still unclear how to implement these methods
to allow for quantum side information. In this section we will show that it is possible to generalize all
uncertainty relations for the sum of Shannon entropies to allow for quantum side information by using the {\it Holevo inequality}.

Before analyzing our main techniques and results, let us first discuss the modern
formulation of the uncertainty principle, the so-called {\it guessing game} (also known as the {\it uncertainty game}),
which highlights its relevance with quantum cryptography. We can imagine there are two observers,
Alice and Bob. Before the game initiates, they agree on two measurements $M_{1}$ and $M_{2}$.
The guessing game proceeds as follows: Bob, can prepare an arbitrary state $\rho_{A}$ which he will send to Alice.
Alice then randomly chooses to perform one of measurements and records the outcome.
After telling Bob the choices of her measurements, Bob can win the game if he correctly guesses Alice's outcome.
Nevertheless, the uncertainty principle tells us that Bob cannot win the game under the condition of incompatible measurements.

What if Bob prepares a bipartite quantum state $\rho_{AB}$ and sends only the particle $A$ to Alice? Equivalently,
what if Bob has nontrivial {\it quantum side information} about Alice's system? Or, what if all information Bob has on
the particle $\rho_{A}$ is beyond the classical description, for example, information on its density matrix?
Berta {\it et al.} \cite{B} answered these questions and generalized the uncertainty relation  (\ref{MU}) to
the case with an auxiliary quantum system $B$ known as quantum memory.

It is now possible for Bob to experience no uncertainty at all when equipped himself with quantum memory, and Bob's uncertainty
about the result of measurements on Alice's system is bounded by
\begin{align}\label{Berta}
H(M_{1}|B)+H(M_{2}|B)\geqslant\log_2\frac{1}{c_{1}}+H(A|B),
\end{align}
where $H(M_{1}|B)=H(\rho_{M_{1}B})-H(\rho_{B})$ is the conditional entropy with
$\rho_{M_{1}B}=\sum_{j}(|u_{j}\rangle\langle u_{j}|\otimes I)(\rho_{AB})(|u_{j}\rangle\langle u_{j}|\otimes I)$ (similarly for $H(M_{2}|B)$),
and the term $H(A|B)=H(\rho_{AB})-H(\rho_{B})$ is related to the entanglement between
 the measured particle $A$ and the quantum memory $B$.

On the other hand, entropic uncertainty relation without quantum memory can be roughly divided into two categories. If the
measure of incompatibility is effective (state-dependent), one can follow Kaniewski, Tomamichel and Wehner's approach to obtain bounds
(e.g. $B_{ac}$ \cite{Kaniewski}) based on effective anticommutators. Otherwise one can derive strong bounds (e.g. $B_{Maj1}$,
$B_{Maj2}$, $B_{RPZ1}$, $B_{RPZ2}$, $B_{RPZ3}$ \cite{Rudnicki}) based on majorization, or bounds (e.g. $B_{CP}$ \cite{Coles}) constructed by the
monotonicity of relative entropy under quantum chan\-nels. Note that Maassen and Uffink's bound $B_{MU}$ \cite{Maassen}, Coles and
Piani's bound $B_{CP}$ \cite{Coles} are still valid in the presence of quantum memory by adding an extra term $H(A|B)$. All these
bounds can be generalized to allow for quantum side information.

Suppose we are given a quantum state $\rho_{AB}$ and a pair of observables, $M_{m}$ ($m=1, 2$). Define the {\it classical
correlation} of state $\rho_{AB}$ with respect to the measurement $M_{m}$ by
\begin{align}
H(\rho_{B})-S_{m}
\end{align}
with
 $$S_{m}=\sum\limits_{i_{m}}p^{m}_{i_{m}}H(\rho^{m}_{B_{i_{m}}}),$$
where $\rho_{B_{i_{m}}}^{m}=Tr_{A}(|u^{m}_{i_{m}}\rangle\langle u^{m}_{i_{m}}|\rho_{AB})/p^{m}_{i_{m}}$ and
$(p^{m}_{i_{m}})_{i_{m}}$ is the probability vector according to the measurement $M_{m}$.

It follows from definition and {\it Holevo's inequality} \cite{Nielsen} that the entropic uncertainty relation in the presence
of quantum memory can be written as
\begin{align}
H(M_{1}|B)+H(M_{2}|B)
=H(M_{1})+H(M_{2})-2H(\rho_{B})+S_{1}+S_{2},
\end{align}
where $H(M_{1}), H(M_{2})$ are the Shannon entropies of the state $\rho_{A}$. Suppose
$B_C$ is a lower bound of the entropic sum $H(M_1)+H(M_2)$, then %it is clearly $B_{C}$ is part of the bound for $H(M_{1}|B)+H(M_{2}|B)$
\begin{align}\label{e:quantum measure}
H(M_{1}|B)+H(M_{2}|B)
\geqslant B_{C}-2H(B)+S_{1}+S_{2}.
\end{align}
We analyze the lower bound according to various types of $B_C$ as follows. In Table 1, we list the various bounds such
as $B_{MU}$, $B_{CP}$, etc. and their
references.

(\expandafter{\romannumeral1}) Bounds \cite{Maassen, Coles, Rudnicki}
that contain a nonnegative state-dependent term $H(A)=S(\rho_{A})$, the {\it von Neumann} entropy
({\it mixing part}):
\begin{align}
H(M_{1})+H(M_{2})&\geqslant B_{MU}+H(A);\notag\\
H(M_{1})+H(M_{2})&\geqslant B_{CP}+H(A);\notag\\
H(M_{1})+H(M_{2})&\geqslant B_{RPZm}+H(A). ~(m=1,2,3)
\end{align}

(\expandafter{\romannumeral2}) Bounds \cite{Kaniewski, Rudnicki, Friedland, Puchala} without the mixing term $H(A)$:
\begin{align}
H(M_{1})+H(M_{2})&\geqslant B_{ac},\notag\\
H(M_{1})+H(M_{2})&\geqslant B_{Majm}. ~(m=1,2)
\end{align}

\begin{figure}
\centering
\includegraphics[width=0.45\textwidth]{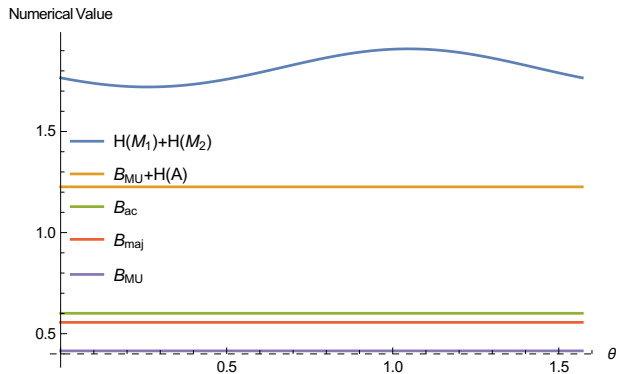}
\caption{ Comparison of bounds for the quantum state $\rho_{A}$ in (\ref{rho1}). The blue, orange, green, red and purple curves are
 respectively the entropic sum $H(M_{1})+H(M_{2})$, the entropic bound $B_{MU}+H(A)$, the entropic bound $B_{ac}$, the entropic bound
 $B_{maj}$, and Maassen and Uffink's bound $B_{MU}$.  The dashed line is $\theta$-axis.}
\end{figure}

Although both effective anticommutators and majorization approach play an important role in improving the bound for entropic
uncertainty relations, even the strengthened Maassen and Uffink's bound $B_{MU}+H(A)$ can
be tighter than the majorization bound $B_{Maj1}$ \cite{Rudnicki} and Kaniewski-Tomamichel-Wehner's bound $B_{ac}$
\cite{Kaniewski} if the mixing part is absent. To see
this, we consider a family of quantum states
\begin{align}\label{rho1}
\rho_{A}=\frac{1}{2}
&\left(
\begin{array}{cc}
  \cos^{2}\theta+\frac{1}{2} & \cos\theta\sin\theta \\
  \cos\theta\sin\theta & \sin^{2}\theta+\frac{1}{2}
\end{array}
\right),
\end{align}
where $0\leqslant\theta\leqslant\pi/2$ with the measurements
$M_{1}=\{(1,0),(0,1)\}$ and $M_{2}=\{(1/2,-\sqrt{3}/2),(\sqrt{3}/2,1/2)\}$. The relations among $H(M_{1})+H(M_{2})$,
$B_{Maj}$ \cite{Rudnicki}, $B_{ac}$ \cite{Kaniewski}, $B_{MU}$ \cite{Maassen} and $B_{MU}+H(A)$ are shown in FIG. 1.
The maximum overlap is $c_{1}=3/4$, and it is known \cite{Kaniewski}
that the bound $B_{ac}$ outperforms $B_{Maj}$. Moreover, the picture shows that the quantity $B_{MU}+H(A)$ is tighter than either
$B_{Maj}$ or $B_{ac}$.

In the above discussion the value $H(A)$ is a constant, so all the bounds
appeared in FIG. 1 are straight lines. Now let's turn to the quantum
states given by
\begin{align}\label{rho2}
\rho_{A}=\frac{1}{2}
&\left(
\begin{array}{cc}
  \cos^{2}\theta & 0 \\
  0 & \sin^{2}\theta
\end{array}
\right),
\end{align}
where $0\leqslant\theta\leqslant\pi/2$ with the same measurements as above. The
relations among $H(M_{1})+H(M_{2})$, $B_{Maj}$, $B_{ac}$, $B_{MU}$ and $B_{MU}+H(A)$ are depicted
in FIG. 2, once again the strengthened Maassen-Uffink's bound $B_{MU}+H(A)$ outperforms both $B_{Maj}$ and $B_{ac}$. In the
neighborhood of $\theta=\pi/4$, the bound $B_{MU}+H(A)$ gives the best estimate.

\begin{figure}
\centering
\includegraphics[width=0.45\textwidth]{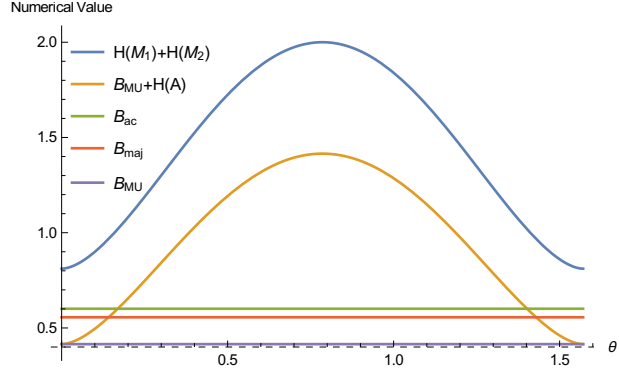}
\caption{Comparison of bounds for quantum state $\rho_{A}$ from  (\ref{rho2}). The blue, orange, green, red, and purple curves are respectively the entropic bounds $H(M_{1})+H(M_{2})$, $B_{MU}+H(A)$, $B_{ac}$,
 $B_{maj}$, and Maassen and Uffink's bound $B_{MU}$. The dashed line is $\theta$-axis.}
\end{figure}

\section{Quantum Measures}

The existence of quantum memory translates into additional information on the uncertainty relation. We introduce the notion
of {\it quantum measure} to describe the relationship between measured particle and quantum memory. There are two types
of quantum measures. 

The first type of quantum measure on entropic uncertainty relations is the mutual information between measured particle $A$ and
quantum memory $B$, which comes from the conditional von Neumann entropy \cite{B}
\begin{align}
H(A|B)=H(A)-I(A:B)
\end{align}
with $I(A:B)=H(A)+H(B)-H(A, B)$ and $H(A, B)=H(\rho_{AB})$. Let $Q_{1}=-I(A:B)$ be the {\it first quantum measure}, as $H(A)$ counts for the mixing level for
measured particle $A$.  Then the bounds for the entropic uncertainty relation in the presence of quantum memory consist of three parts:
the bound $B_{C}$ for the sum of Shannon entropies, the {\it mixing part} $H(A)$ and the first quantum measure $Q_{1}$
\begin{align}
H(M_{1}|B)+H(M_{2}|B)&\geqslant B_{MU}+H(A)+Q_{1},\notag\\
H(M_{1}|B)+H(M_{2}|B)&\geqslant B_{CP}+H(A)+Q_{1},
\end{align}
where $B_{MU}=-\log c_{1}$, $B_{CP}=-\log c_{1}+\frac{1-\sqrt{c_{1}}}{2}\log\frac{c_{1}}{c_{2}}$, and $c_{2}$ is the second
largest entry of the matrix $(\mid\langle u^{1}_{i_{1}}|u^{2}_{i_{2}}\rangle\mid^{2})_{i_{1}i_{2}}$.

A more natural and less restrictive {\it quantum measure} is $-2H(B)+S_{1}+S_{2}$ discussed in Sec.
\uppercase\expandafter{\romannumeral2}. Let $Q_{2}=-2H(B)+S_{1}+S_{2}$ be the {\it second quantum
measure}, then we can generalize all the bounds for the sum of Shannon entropies to allow
for quantum side information. Namely we have
\begin{align}
H(M_{1}|B)+H(M_{2}|B)&\geqslant B_{MU}+H(A)+Q_{2},\notag\\
H(M_{1}|B)+H(M_{2}|B)&\geqslant B_{CP}+H(A)+Q_{2},
\end{align}

Clearly, both Maassen and Uffink's bound $B_{MU}$ and Coles and Piani's bound $B_{CP}$ are valid with or without
quantum side information, with the mixing part $H(A)$ in the former case or the conditional entropy $H(A|B)$ in the latter.
Mathematically, the relation says that
\begin{align}\label{e:bound1}
H(M_{1})+H(M_{2})&\geqslant B_{CC}+H(A),\notag\\
H(M_{1}|B)+H(M_{2}|B)&\geqslant B_{CC}+H(A|B),
\end{align}
where $B_{CC}=B_{MU}$ or $B_{CP}$. The term $B_{CC}$ will be referred as the {\it consistent classical part} of the bound for
the entropic uncertainty relation.
In place of $B_{MU}$ and $B_{CP}$ in \eqref{e:bound1}, we have recently given a new consistent classical part $B$, which is a
tighter bound
depending on all overlaps between incompatible observables
\cite{XJ}:
\begin{align}\label{d}
B=\log_2 \frac1{c_{1}}+\frac{1-\sqrt{c_{1}}}{2}\log_2\frac{c_{1}}{c_{2}}+\frac{2-\Omega_{4}}{2}\log_2\frac{c_{2}}{c_{3}}+
%\frac{2-\Omega_{6}}{2}\log_2\frac{c_{3}}{c_{4}}
\cdots+\frac{2-\Omega_{2(d-1)}}{2}\log_2\frac{c_{d-1}}{c_{d}},
\end{align}
where $c_{i}$ is the $i$-th largest overlap among $c_{jk}$: $c_{1}\geqslant c_{2}\geqslant c_{3}\geqslant\cdots\geqslant c_{d^{2}}$,
and $\Omega_{k}$ is the $k$-th element of majorization bound for measurements $M_{1}$ and $M_{2}$ \cite{XJ}.
In general the bound $B$ is always tighter than $B_{CP}$, except possibly when two orthonormal bases are mutually unbiased.

\begin{table*}[ht]
\caption{Comparison among bounds for entropic uncertainty relations with and without quantum memory} % title of Table
\centering % used for centering table
\begin{tabular}{c | c | c} % centered columns (4 columns)
\hline\hline %inserts double horizontal lines
Reference &
Lower bound for $H(M_{1})+H(M_{2})$ &
Lower bound for $H(M_{1}|B)+H(M_{2}|B)$ \\ [0.5ex] % inserts table
%heading
\hline % inserts single horizontal line
\cite{Maassen} & $B_{MU}+H(A)$ & $B_{MU}+H(A)+Q_{1}$ (or $Q_{2}$) \\ % inserting body of the table
\hline
\cite{Coles} & $B_{CP}+H(A)$ & $B_{CP}+H(A)+Q_{1}$ (or $Q_{2}$) \\
\hline
\cite{XJ} & $B+H(A)$ & $B+H(A)+Q_{1}$ (or $Q_{2}$) \\
\hline
\cite{Kaniewski} & $B_{ac}$ & $B_{ac}+Q_{2}$ \\
\hline
\cite{Rudnicki} & $B_{Maj1}$ & $B_{Maj1}+Q_{2}$ \\
\hline
\cite{Rudnicki} & $B_{Maj2}$ & $B_{Maj2}+Q_{2}$ \\
\hline
\cite{Rudnicki} & $B_{RPZ1}+H(A)$ & $B_{RPZ1}+H(A)+Q_{2}$ \\
\hline
\cite{Rudnicki} & $B_{RPZ2}+H(A)$ & $B_{RPZ2}+H(A)+Q_{2}$ \\
\hline
\cite{Rudnicki} & $B_{RPZ3}+H(A)$ & $B_{RPZ3}+H(A)+Q_{2}$ \\ [1ex] % [1ex] adds vertical space
\hline %inserts single line
\end{tabular}
\label{table:quantum memory} % is used to refer this table in the text
\end{table*}

We continue discussing the quantum measure of the entropic uncertainty relation with a consistent classical part.
When quantum memory is present, there are infinitely many
quantum measures. For any $\lambda\in[0,1]$ one has that
\begin{align}
H(M_{1}|B)+H(M_{2}|B)\geqslant B_{CC}+H(A)+Q(\lambda),
\end{align}
where
\begin{align}
Q(\lambda):=-\lambda I(A:B)+(1-\lambda)(-2H(B)+S_{1}+S_{2})
\end{align}
is a new quantum measure for the entropic uncertainty relation with a consistent part.
Here we have used a weighted sum of quantum measures similar to \cite{Weighted}. Note that the weight is
applied on the quantum measures instead of the uncertainty relations. Through this simple process, we can always
get a better lower bound without worrying which quantum measure is tighter than the other.
Aside from its own significance, the new quantum measure $Q(\lambda)$ is expected to be useful for
future quantum technologies such as entanglement witnessing.

The quantum measure $Q_{2}$ has two desirable features. First, with the help of the {\it second
quantum measure} we can extend all previous bounds of the entropic sum (Shannon entropy) to allow for the quantum
side information without restrictive constraints. The comparison of some of the existing results is given together
with their extensions in the presence of quantum side information in TABLE. 1. Second, $Q_{2}$ can sometimes outperform
$Q_{1}$ to give tighter bounds for the entropic uncertainty relation in the presence of quantum memory. For more details, see
Sec. \uppercase\expandafter{\romannumeral4}.

Third, by taking the maximum over $Q_{2}-Q_{1}$ and zero, we derive that
\begin{align}
\max\{0, Q_{2}-Q_{1}\},
\end{align}
is another bound, i.e. $H(M_{1}|B)+H(M_{2}|B)\geqslant B_{C}+H(A|B)+\max\{0, Q_{2}-Q_{1}\}$, which coincides with the main quantity used in the recent paper \cite[Eq (12)]{Adabi}
for a strong uncertainty relation in the presence of quantum
memory. We point it out that our result is more general than simply using $\max\{0, Q_{2}-Q_{1}\}$. In fact, $B+H(A)+\max\{Q_{1}, Q_{2}\}$ is tighter than the outcomes from \cite{Adabi}. In \cite{XJ} we have given a detailed and rigourous proof on
the lower bound.

\section{Influence of Incompatible Observables}

Let us consider two pairs of incompatible observables $M_{1}$, $M_{2}$
and $M_{3}$, $M_{4}$ with the same overlaps $c_{jk}$. Then the bounds for the Shannon entropic sum $H(M_{1})+H(M_{2})$ on measured particle
$A$ will coincide with that of $H(M_{3})+H(M_{4})$, since their bounds only depend on the overlaps $c_{jk}$.
If there is quantum memory $B$ present, the same relation holds for the bounds with the first quantum measure $Q_1$, since their bounds also  depend
only on $c_{jk}$ and $H(A|B)$.

However,
the situation is quite different by utilizing the {\it second quantum measure}. Even when two pairs of incompatible observables $M_{1}$,
$M_{2}$ and $M_{3}$, $M_{4}$ share the same overlaps, the corresponding bounds may differ. This interesting phenomenon
may be useful in physical experiments: the total uncertainty can be decreased by choosing suitable incompatible observables.

As an example, consider the following
$2\times 4$ bipartite state,  %$\rho_{AB}$
\begin{align}\label{e:state1}
\rho_{AB}=\frac{1}{1+7p}
&\left(
\begin{array}{cccccccc}
  p & 0 & 0 & 0 & 0 & p & 0 & 0 \\
  0 & p & 0 & 0 & 0 & 0 & p & 0 \\
  0 & 0 & p & 0 & 0 & 0 & 0 & p \\
  0 & 0 & 0 & p & 0 & 0 & 0 & 0 \\
  0 & 0 & 0 & 0 & \frac{1+p}{2} & 0 & 0 & \frac{\sqrt{1-p^2}}{2} \\
  p & 0 & 0 & 0 & 0 & p & 0 & 0 \\
  0 & p & 0 & 0 & 0 & 0 & p & 0 \\
  0 & 0 & p & 0 & \frac{\sqrt{1-p^2}}{2} & 0 & 0 & \frac{1+p}{2}
\end{array}
\right),
\end{align}
which is known to be entangled for $0<p<1$ \cite{H}. We take system $A$ as the quantum memory and measurements are performed on
system $B$. Choose the incompatible observables $M_{1}=\{|u^{1}_i\rangle\}$ and $M_{2}=\{|u^{2}_i\rangle\}$ as the first
pair of measurements
\begin{align}\label{pair}
&|u^1_1\rangle=(\frac{1}{\sqrt{2}},-\frac{1}{\sqrt{2}},0,0)^\dag,|u^1_2\rangle=(\frac{1}{\sqrt{2}},\frac{1}{\sqrt{2}},0,0)^\dag,\notag\\
&|u^1_3\rangle=(0,0,\frac{1}{\sqrt{2}},\frac{1}{\sqrt{2}})^\dag,|u^1_4\rangle=(0,0,\frac{1}{\sqrt{2}},-\frac{1}{\sqrt{2}})^\dag;\notag\\
&|u^2_1\rangle=\frac{1}{\sqrt{6}}(\sqrt{2},\sqrt{2},\sqrt{2},0)^\dag,
|u^2_2\rangle=\frac{1}{\sqrt{6}}(\sqrt{3},0,-\sqrt{3},0)^\dag,\notag\\
&|u^2_3\rangle=\frac{1}{\sqrt{6}}(1,-2,1,0)^\dag,
|u^2_4\rangle=(0,0,0,1)^\dag,
\end{align}
then take $M_3=M_2$ and $M_{4}=\{|u^3_i\rangle\}$ such that
\begin{align}
|u^1_j\rangle&\neq|u^3_j\rangle,\notag\\
\mid\langle u^2_j|u^3_k\rangle\mid^{2}&=\mid\langle u^1_j|u^2_k\rangle\mid^{2}.
\end{align}
Therefore, the basis $M_{4}$ is obtained as
\begin{align}
(|u^2_1\rangle, |u^2_2\rangle, |u^2_3\rangle, |u^2_4\rangle)&=U(|u^1_1\rangle, |u^1_2\rangle, |u^1_3\rangle, |u^1_4\rangle),\notag\\
(|u^3_1\rangle, |u^3_2\rangle, |u^3_3\rangle, |u^3_4\rangle)&=U(|u^2_1\rangle, |u^2_2\rangle, |u^2_3\rangle, |u^2_4\rangle),
\end{align}
where the matrix $U$ is easily fixed from  (\ref{pair}).

\begin{figure}
\centering
\includegraphics[width=0.45\textwidth]{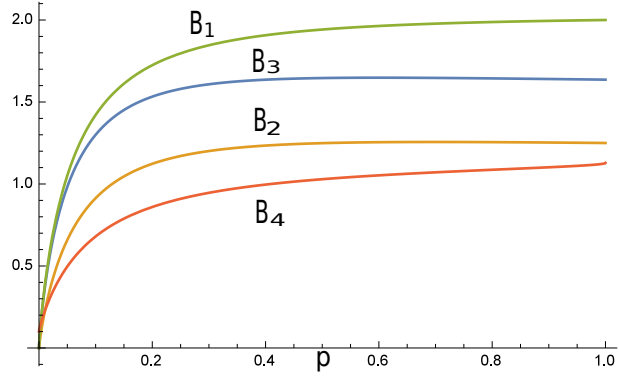}
\caption{Comparison of bounds for entangled quantum state $\rho_{AB}$. The green curve is the entropic bound $\mathcal{B}_1$,
 the blue curve is the entropic bound $\mathcal{B}_3$, the orange curve is the entropic bound $\mathcal{B}_2$
 and the red curve is the entropic bound $\mathcal{B}_4$.}
\end{figure}

Set $\mathcal{B}_1=H(B)$, $\mathcal{B}_2=H(B|A)$, $\mathcal{B}_3=H(B)-2H(A)+S_1+S_2$, $\mathcal{B}_4=H(B)-2H(A)+S_2+S_3$
and $\mathcal{B}_c:=B$ (cf. (\ref{d})). If there is no quantum memory,
the entropic uncertainty relations are obtained as
\begin{align}
&H(M_{1})+H(M_{2})\geqslant\mathcal{B}_c+\mathcal{B}_1,\notag\\
&H(M_{3})+H(M_{4})\geqslant\mathcal{B}_c+\mathcal{B}_1,
\end{align}
where the bounds are the same due to identical overlaps between the bases.
In the presence of quantum memory, using the {\it first quantum measure}  $Q_{1}$ as the extra term to describe the amount of
correlations between measured particle and quantum memory, we have that
\begin{align}
&H(M_{1}|A)+H(M_{2}|A)\geqslant\mathcal{B}_c+\mathcal{B}_2,\notag\\
&H(M_{3}|A)+H(M_{4}|A)\geqslant\mathcal{B}_c+\mathcal{B}_2,
\end{align}
so their bounds coincide again.
Finally, choosing the {\it second quantum measure} $Q_{2}$ for the correlations between measured particle and
quantum memory, we derive that
\begin{align}
&H(M_{1}|A)+H(M_{2}|A)\geqslant\mathcal{B}_c+\mathcal{B}_3,\notag\\
&H(M_{3}|A)+H(M_{4}|A)\geqslant\mathcal{B}_c+\mathcal{B}_4,
\end{align}
and this time their bounds are different from each other. Therefore when the measured particle and quantum memory are entangled,
the uncertainty is decreased through suitable incompatible observables. Since all the
bounds contain  $\mathcal{B}_c$,
we only need to compare $\mathcal{B}_1=H(B)$, $\mathcal{B}_2=H(B|A)$,
$\mathcal{B}_3=H(B)-2H(A)+S_1+S_2$ and $\mathcal{B}_4=H(B)-2H(A)+S_2+S_3$ for two
pairs of measurements.

In FIG. 3, the comparison is done for $\mathcal{B}_1$, $\mathcal{B}_2$, $\mathcal{B}_3$
and $\mathcal{B}_4$, which shows how the {\it second quantum measure} works for selected pairs of incompatible observables. The bound $\mathcal{B}_3$
(with the {\it second quantum measure}) provides the best estimation for the entropic sum in the presence of quantum memory, while the bound
$\mathcal{B}_2$ (with the {\it first quantum measure}) gives a weaker approximation. The {\it second quantum measure} does not always outperform
the {\it first quantum measure}, since $\mathcal{B}_4$ is typically worse than $\mathcal{B}_2$. However, comparing the bound $\mathcal{B}_3$ with
$\mathcal{B}_4$, we find that the uncertainty from measurements can be weaken by selecting appropriate measurements even if each pair of incompatible
observables shares the same overlaps.

To illustrate improvement of the bound in the presence of quantum memory, we compare the bound based on the
{\it second quantum measure} with that based on the {\it first quantum measure}. As a first step, choose the initial state
as {\it Werner State} $\rho_{AB}=\frac{1}{4}(1-p)I+p|B_{1}\rangle\langle B_{1}|$ with $0<p<1$, and
$|B_{1}\rangle=\frac{1}{\sqrt{2}}(|00\rangle+|11\rangle)$ the {\it Bell State}. Take
$|u^1_1\rangle=(\frac{1}{\sqrt{2}},-\frac{1}{\sqrt{2}})$, $|u^1_2\rangle=(\frac{1}{\sqrt{2}},\frac{1}{\sqrt{2}})$;
$|u^2_1\rangle=(\cos\theta,-\sin\theta)$, $|u^2_2\rangle=(\sin\theta,\cos\theta)$ with $0<\theta<2\pi$, then the difference between the
bound with {\it second quantum measure} and the bound with the {\it first quantum measure} is illustrated in  FIG. 4. The nonnegativity
of the surface shows that our newly constructed bound with the {\it second quantum measure} can outperform the bound with the {\it first quantum measure} everywhere in this case.

\begin{figure}
\centering
\includegraphics[width=0.45\textwidth]{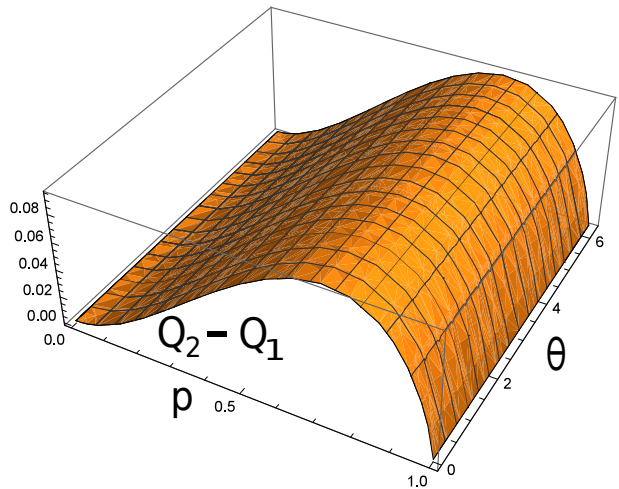}
\caption{The difference between the bound of entropic uncertainty relations in the presence of quantum memory with the {\it second quantum measure}
$Q_{2}$ and the bound of entropic uncertainty relations in the presence of quantum memory with the {\it first quantum measure}
$Q_{1}$.}
\end{figure}

Using quantum measures we have shown that it is possible to reduce the total uncertainties coming from
incompatibility of the observables by an appropriate choice. However, when the measured particle and quantum memory are maximally entangled,
both the first and second {\it quantum measure} equal to $-\log_{2}d$. We sketch a proof of this statement in Appendix A.

\section{Strong Entropic Uncertainty Relations in the presence of quantum memory}

In this section, we derive several strong entropic uncertainty relations in the presence of quantum memory by utilizing both the relevant
bounds for the sum of Shannon entropies and optimal selection of quantum measures. Recall that the bounds of entropic uncertainty
relations in the presence of quantum memory contain three ingredients: the classical part $B_{C}$, the mixing part $H(A)$ (which is not necessarily existent,
e.g., the majorization bounds \cite{Rudnicki, Friedland, Puchala} and $B_{ac}$ \cite{Kaniewski}), and the quantum measures $Q_{i}$ ($i=1,2$).

Let $\rho_{AB}$ be a bipartite quantum state, and $M_{i}$ ($i=1, 2$) two nondegenerate incompatible observables
on the system $A$. We take system $B$ as the quantum memory. A simple lower bound for the entropic sum
in the presence of quantum memory can be obtained as follows. Note that the {\it consistent classical part} $B_{CC}$
is valid with both quantum measures $Q_{i}$, therefore for $i=1, 2$
\begin{align}
H(M_{1}|B)+H(M_{2}|B)\geqslant&B_{CC}+H(A)+Q_{i}.
\end{align}

As the bound $B$ in (\ref{d}) is the tightest, %{\it consistent classical part} till now,
so the strongest lower bound for the entropic sum
in the presence of quantum memory with {\it consistent classical part} is given by
\begin{align}\label{e:cc}
\mathfrak{B}_{CC}:=B+H(A)+\max\{Q_{1},Q_{2}\}.
\end{align}

Without the help of the consistent classical part, all other classical parts $B_{C}$ can be estimated in the
same way.
\begin{align}\label{e:secondindex}
H(M_{1}|B)+H(M_{2}|B)\geqslant\mathcal{B}_{C}+H(A)+Q_{2}.
\end{align}
Note that for $B_{C}=B_{ac}$ or $B_{Maj}$, there is no mixing part $H(A)$ on the right-hand side
of  (\ref{e:secondindex}). Taking the maximum over all possible $B_{C}$'s %of the right-hand side of  (\ref{e:secondindex}),
we obtain a lower bound %denote by $\mathfrak{B}_{C}$
\begin{align}
\mathfrak{B}_{C}:=\max\{B_{ac}, B_{Maj1}, B_{Maj2}, B_{RPZ1}+H(A),
B_{RPZ2}+H(A), B_{RPZ3}+H(A)\}+Q_{2},
\end{align}

Clearly both the lower bounds $\mathfrak{B}_{C}$ and $\mathfrak{B}_{CC}$ can be combined into a hybrid bound for
the uncertainty relation in the presence of quantum memory: %hybrid
%for the sum of entropies
\begin{align}\label{biparticle}
H(M_{1}|B)+H(M_{2}|B)\geqslant\max\{\mathfrak{B}_{C},\mathfrak{B}_{CC}\},
\end{align}
where $\mathfrak{B}_{C}$ and $\mathfrak{B}_{CC}$ are given by (\ref{e:cc}) and (\ref{e:secondindex}) respectively.

We now extend our results to the general case of $L$-partite particles ($L\geqslant3$) with $N$ incompatible
observables ($N\geqslant3$). Assume the measured system is the $l_{1}$-partite subsystem and the
quantum memory is the remaining $l_{2}$-partite subsystem, where $l_{2}=L-l_1$ and $l_{1}\geqslant2$.

Suppose that the $N$ measurements $M_{1}$, $M_{2}$, $\ldots$, $M_{N}$ are given by the bases $M_{m}=\{|u^{m}_{i_{m}}\rangle\}$.
Let system $A$ be the measured
particle ($l_{1}$-partite) and $B$ the quantum memory ($l_{2}$-partite). The probability distributions
$$p^{m}_{i_{m}}=\langle u^{m}_{i_{m}}|\rho_{A}|u^{m}_{i_{m}}\rangle,$$
have a majorization bound  \cite{Hossein}: % $\omega=\sup\limits_{M_{m}}(p^{m}_{i_{m}})$
\begin{align}
(p^{m}_{i_{m}})\prec \omega=\sup\limits_{M_{m}}(p^{m}_{i_{m}}),
\end{align}
which is state-independent. For different correlations between particles, there may exist different kind of state-independent $\omega$ called
 the {\it uniform entanglement frames} \cite{XiaoJing}. % for more detail see \cite{XiaoJing}.
 In fact, if the majorization bound is written as
 $\omega=(\Omega_{1},\Omega_{2}-\Omega_{1},\cdots,1-\Omega_{d-1})$, then we have
\begin{align}\label{e:multi}
\sum\limits_{m=1}^{N}H(M_{m}|B)\geqslant(N-1)H(A|B)-\log_2 b_{1}
+(1-\Omega_{1})\log_2\frac{b_{1}}{b_{2}}+\cdots+(1-\Omega_{d-1})\log_2\frac{b_{d-1}}{b_{d}},
\end{align}
where $b_{i}$ is the $i$-th largest element among all
$$\left\{\sum\limits_{i_{2}\cdots i_{N-1}}\max_{i_{1}}[c(u^{1}_{i_{1}},u^{2}_{i_{2}})]
\prod\limits_{m=2}^{N-1}c(u^{m}_{i_{m}},u^{m+1}_{i_{m+1}})\right\}$$
over the indices $i_{N}$ and $c(u^{m}_{i_{m}},u^{m+1}_{i_{m+1}})=\mid\langle u^{m}_{i_{m}}|u^{m+1}_{i_{m+1}}\rangle\mid^{2}$.
A complete proof of the relation \eqref{e:multi} is given in Appendix B. %We emphasize that our bound (\ref{e:multi})

Besides giving theoretical improvement of the uncertainty relation,
our result has potential applications in other areas of quantum theory. For example,
it can be utilized in designing new entanglement detector. To witness entanglement, one considers
a source that emits a bipartite state $\rho_{A}$. One defines the probability distributions of incompatible observables
$M_{m}$ ($m=1, \cdots, N$) as usual:
$$p^{m}_{i_{m}}=\langle u^{m}_{i_{m}}|\rho_{A}|u^{m}_{i_{m}}\rangle.$$
If the bipartite state $\rho_{A}$ is separable, then there exists a vector
$\omega^{sep}=(\Omega_{1}^{sep},\Omega_{2}^{sep}-\Omega_{1}^{sep},\cdots,1-\Omega_{d-1}^{sep})$ such that
\begin{align}
(p^{m}_{i_{m}})\prec \omega^{sep}.
\end{align}
Subsequently we have
\begin{align}
\sum\limits_{m=1}^{N}H(M_{m})\geqslant(N-1)H(A)-\log_2 b_{1}
+(1-\Omega_{1}^{sep})\log_2\frac{b_{1}}{b_{2}}+\cdots+(1-\Omega_{d-1}^{sep})\log_2\frac{b_{d-1}}{b_{d}},
\end{align}
with other notations are the same with  (\ref{e:multi}). If there exists another quantum state $\rho_{A}^{\prime}$ with
\begin{align}
\sum\limits_{m=1}^{N}H(M_{m})<(N-1)H(A^{\prime})-\log_2 b_{1}
+(1-\Omega_{1}^{sep})\log_2\frac{b_{1}}{b_{2}}+\cdots+(1-\Omega_{d-1}^{sep})\log_2\frac{b_{d-1}}{b_{d}},
\end{align}
where $H(A^{\prime})=S(\rho_{A}^{\prime})$, then state $\rho_{A}^{\prime}$ must be entangled since it violates the majorization
bound for separable states. As this method is based on {\it uniform entanglement frames} and the entropic uncertainty relations,
the witnessed entanglement does not involve with quantum memory.

Similarly, the second quantum measure enables us to generalize the strong entropic uncertainty relations for
multiple measurements \cite{XiaoJingPRA} (i.e. {\it admixture bound}) to allow for quantum side information. By taking the maximum
over (\ref{e:multi}) and the {\it admixture bound} in the presence of quantum memory, we obtain a strong entropic uncertainty
relation with quantum memory for multi-measurements which will be useful in handling quantum cryptography tasks and general quantum
information processings.

\section{Conclusions}

We have extended all uncertainty relations for Shannon entropies to allow for quantum side information, first in the case of
two incompatible observables and then for multi-observables. Using the {\it second quantum measure} we have characterized the correlations
between measured particle and quantum memory. Our uncertainty relations are universal and capture the intrinsic nature of
the uncertainty in the presence of quantum memory. Moreover, we have observed that the uncertainties in the presence of quantum memory
decrease under appropriate selection of incompatible observables. Finally, we have derived several strong bounds for the entropic uncertainty
relation in the presence of quantum memory. We have also discussed applications of our result to entanglement witnesses with or without
quantum memory.

\medskip
\noindent{\bf Acknowledgments}\, \,
We thank Shao-Ming Fei for interesting discussions on related topics.
The work is supported by National Natural Science Foundation of China (grant Nos. 11271138 and 11531004), 
China Scholarship Council and Simons Foundation (grant no. 198129).

\section{Appendix A: Maximal Entanglement}

Let $\rho_{AB}$ be a bipartite quantum state, and $M_{1}$, $M_{2}$ a pair of incompatible observables. Suppose that
the measured particle $A$ and quantum memory $B$ are maximally entangled. We will show that
both the first and second quantum measures coincide with each other. Recall that the first
quantum measure $Q_{1}$ was defined in Sec. \uppercase\expandafter{\romannumeral3} and the combination of the quantum measure
and mixing part is $H(A)+Q_{1}=H(A|B)=-\log_2 d$.

Recall that the second quantum measure is given by $Q_{2}=-2H(B)+S_{1}+S_{2}$, where %let us write the terms $S_{1}$ and $S_{2}$ as follows:
\begin{align}
S_{1}&=\sum\limits_{i_{1}}p^{1}_{i_{1}}H(\rho^{1}_{B_{i_{1}}}),\\
S_{2}&=\sum\limits_{i_{2}}p^{2}_{i_{2}}H(\rho^{2}_{B_{i_{2}}}).
\end{align}
From $p^{m}_{i_{m}}=\langle u^{m}_{i_{m}}|\rho_{A}| u^{m}_{i_{m}}\rangle$ and
$[u^{m}_{i_{m}}]\equiv|u^{m}_{i_{m}}\rangle\langle u^{m}_{i_{m}}|$ ($m=1, 2$), it follows that
\begin{align}
\rho^{1}_{B_{i_{1}}}&=\frac{Tr_{A}([u^{1}_{i_{1}}]\rho_{AB})}{p^{1}_{i_{1}}},\\
\rho^{2}_{B_{i_{2}}}&=\frac{Tr_{A}([u^{2}_{i_{2}}]\rho_{AB})}{p^{2}_{i_{2}}}.
\end{align}

One can use the formula to compute the second quantum measure $Q_{2}$ if the state is the maximally entangled
quantum state $\rho_{AB}=\frac{1}{\sqrt{d}}\sum\limits_{i=0}^{d-1}|ii\rangle$. For simplicity, we only consider the case
$d=3$ while the high dimensional case can be similarly done. For the projective rank-$1$ measurements
on system $A$, set $|u^{1}_{i_{1}}\rangle=\alpha|0\rangle+\beta|1\rangle+\gamma|2\rangle$
with $\mid\alpha\mid^{2}+\mid\beta\mid^{2}+\mid\gamma\mid^{2}=1$, then
\begin{align}
[u^{1}_{i_{1}}]=
\left(
\begin{array}{ccc}
  \mid\alpha\mid^{2} & \alpha\beta^{\ast} & \alpha\gamma^{\ast} \\
  \beta\alpha^{\ast} & \mid\beta\mid^{2} & \beta\gamma^{\ast} \\
  \gamma\alpha^{\ast} & \gamma\beta^{\ast} & \mid\gamma\mid^{2} \\
\end{array}
\right),
\end{align}
and
\begin{align}
\rho^{1}_{B_{i_{1}}}=
\left(
\begin{array}{ccc}
  \mid\alpha\mid^{2} & \beta\alpha^{\ast} & \gamma\alpha^{\ast} \\
  \alpha\beta^{\ast} & \mid\beta\mid^{2} & \gamma\beta^{\ast} \\
  \alpha\gamma^{\ast} & \beta\gamma^{\ast} & \mid\gamma\mid^{2} \\
\end{array}
\right).
\end{align}

Since the density matrix $\rho^{1}_{B_{i_{1}}}$ is rank $1$, it follows that
\begin{align}
H(\rho^{1}_{B_{i_{1}}})=0,
\end{align}
which implies that $S_{1}=S_{2}=0$. Therefore
$$H(A)+Q_{1}=H(A)+Q_{2}=-\log_2 d,$$
where the last equality implies that the first quantum measure coincide with the second index when the measured
particle and quantum memory are maximally entangled.

\section{Appendix B: Multiple Measurements}

For an $L$-partite state $\rho$, divide the whole system into two parts: the measured subsystem
$A$ and the remaining subsystem as quantum memory $B$, then we can still denote the quantum state as $\rho_{AB}$. Given $N$
measurements $M_{1}, M_{2}, \cdots, M_{N}$, to find a lower bound for the entropic uncertainty relations in the presence of
quantum memory we use basic properties of the relative entropy as follows:
\begin{align}
S(\rho_{AB}\parallel\sum\limits_{i_{1}}[u^{1}_{i_{1}}]\rho_{AB}[u^{1}_{i_{1}}])
\geqslant &S([u^{2}_{i_{2}}]\rho_{AB}[u^{2}_{i_{2}}]\parallel\sum\limits_{i_{1},i_{2}}c(u^{1}_{i_{1}},u^{2}_{i_{2}})
[u^{2}_{i_{2}}]\otimes Tr_{A}([u^{1}_{i_{1}}]\rho_{AB}))\notag\\
=&S(\rho_{AB}\parallel\sum\limits_{i_{1},i_{2}}c(u^{1}_{i_{1}},u^{2}_{i_{2}})
[u^{2}_{i_{2}}]\otimes Tr_{A}([u^{1}_{i_{1}}]\rho_{AB}))
+H(A|B)-H(M_{2}|B),
\end{align}
where $c(u^{1}_{i_{1}}, u^{2}_{i_{2}})=\mid\langle u^{1}_{i_{1}}|u^{2}_{i_{2}}\rangle\mid^{2}$,
$[u^{m}_{i_{m}}]=|u^{m}_{i_{m}}\rangle\langle u^{m}_{i_{m}}|$, and  $S(\rho\parallel\sigma)=\Tr(\rho(\log\rho-\log\sigma))$
stands for the relative entropy.

Inductively the generalized lower bound is given as follows
\begin{align}
-NH(A|B)+\sum\limits_{m=1}^{N}H(M_{m}|B)
\geqslant S(\rho_{AB}\parallel\sum\limits_{i_{N}}[u^{N}_{i_{N}}]\otimes\beta^{N}_{i_{N}}),
\end{align}
where $p^{1}_{i_{1}}\rho^{1}_{B_{i_{1}}}=Tr_{A}([u^{1}_{i_{1}}]\rho_{AB})$ and
$$\beta^{N}_{i_{N}}=\sum\limits_{i_{1},\cdots,i_{N-1}}p^{1}_{i_{1}}\rho^{1}_{B_{i_{1}}}
\prod\limits_{m=1}^{N-1}c(u^{m}_{i_{m}},u^{m+1}_{i_{m+1}})$$
Taking maximum over indices $i_2, \ldots, i_{N-1}$ and writing
\begin{align}
\sum\limits_{i_{2},\cdots,i_{N-1}}\max\limits_{i_{1}}[c(u^{1}_{i_{1}},u^{2}_{i_{2}})]
\prod\limits_{m=2}^{N-1}c(u^{m}_{i_{m}},u^{m+1}_{i_{m+1}})=b(i_{N}),
\end{align}
we have that
\begin{align}
S(\rho_{AB}\parallel\sum\limits_{i_{N}}[u^{N}_{i_{N}}]\otimes\beta^{N}_{i_{N}})
\geqslant-H(A|B)-\sum\limits_{i_{N}}p^{N}_{i_{N}}\log_2 b(i_{N}),
\end{align}
where $p^{N}_{i_{N}}=Tr([u^{N}_{i_{N}}]\rho_{A})$. We arrange the numerical values $b(i_{N})$ in descending
order:
\begin{align}
b_{1}\geqslant b_{2}\geqslant\cdots \geqslant b_{d},
\end{align}
so $b_{i}$ is the $i$-th largest element among all $b(i_{N})$ (counting multiplicity). Denote by $p^{N}_{i}$ the corresponding probability. Therefore
\begin{align}
S(\rho_{AB}\parallel\sum\limits_{i_{N}}[u^{N}_{i_{N}}]\otimes\beta^{N}_{i_{N}})
\geqslant-H(A|B)-\log_2 b_{1}+(1-p_{1})\log_2\frac{b_{1}}{b_{2}}
+\cdots+(1-p_1-\cdots-p_{d-1})\log_2\frac{b_{d-1}}{b_{d}}.
\end{align}

If the measured particle is $l_{1}$-partite and the quantum memory is a $l_{2}$-partite particle such that
$l_{1}+l_{2}=L, l_{1}\geqslant2$, then there exists a state-independent majorization bound \cite{XiaoJing}
$\omega=(\Omega_{1},\Omega_{2}-\Omega_{1},\cdots,1-\Omega_{d-1})$ corresponding to the structure of the measured particle.
Note that
\begin{align*}
1-p_{1}&\geqslant1-\Omega_{1},\\
1-p_{1}-p_{2}&\geqslant1-\Omega_{2},\\
\cdots\cdots\\
1-p_{1}-\cdots-p_{d-1}&\geqslant1-\Omega_{d-1},
\end{align*}
which imply that
\begin{align}
S(\rho_{AB}\parallel\sum\limits_{i_{N}}[u^{N}_{i_{N}}]\otimes\beta^{N}_{i_{N}})
\geqslant-H(A|B)-\log_2 b_{1}
+(1-\Omega_{1})\log_2\frac{b_{1}}{b_{2}}+\cdots+(1-\Omega_{d-1})\log_2\frac{b_{d-1}}{b_{d}}.
\end{align}
Hence the entropic uncertainty relation is written as
\begin{align}\label{e:permutation}
\sum\limits_{m=1}^{N}H(M_{m}|B)\geqslant(N-1)H(A|B)-\log_2 b_{1}
+(1-\Omega_{1})\log_2\frac{b_{1}}{b_{2}}+\cdots+(1-\Omega_{d-1})\log_2\frac{b_{d-1}}{b_{d}},
\end{align}
which provides a substantial improvement over $(N-1)H(A|B)-\log_2 b_{1}$, the term contained in the presence of quantum memory.
Therefore, the new bound is the tightest
one with {\it consistent classical part} till now. By taking all permutations on the index of (\ref{e:permutation}) first,
and computing the maximum over all possibilities, we obtain an optimal lower bound in the presence of quantum memory.
One can also use  uniform entanglement frames \cite{XiaoJing} to give a degenerate uncertainty inequality
in the absence of quantum memory.

\end{document}